**Proposed tests of the soliton wave model of action potentials, and of inducible lipid pores, and how non-electrical phenomena might be consistent with the Hodgkin-Huxley model.**

By:  Scott T. Meissner[1], Prof. (retired)

University of the Philippines Cebu, Department of Biology and Environmental Science, College of Science, Gorordo Avenue, Lahug, Cebu City  6000, The Republic of the Philippines.

**Abstract:**

The soliton wave model of action potentials, and the proposal of induced lipid pores, are potentially paradigm shifting ideas which challenge accepted views of the Hodgkin-Huxley model and of protein-based ion channels.  These two proposals are reviewed, and possible tests of each are presented.  Also, three key non-electrical features seen during action potentials are reviewed;  a shift in birefringence, the pattern of heat emission and absorption, and the expansion of cell diameter.  How the soliton wave model uses the lipid phase transition to account for each of these three phenomena is contrasted with alternatives which might be consistent with the Hodgkin-Huxley model of action potentials.  It is suggested that changes in membrane potential during the action potential might have a significant effect on membrane proteins and contribute to the production of each of these phenomena.  A key assumption of the soliton wave model is that lipid phase transitions are common and adaptive in life.  However a review of the literature suggests that often lipid phase transitions are damaging to cells, and so are often avoided.  There is a need for clearer evidence as to whether or not neurons actually do have liquid crystalline to gel lipid phase transitions happening in them during action potentials, as the soliton wave model assumes.  Several major paradigm shifts are associated with the soliton wave model, and with the proposal of induced lipid pores, and such large claims require additional testing and a great deal of supportive evidence if they are to achieve broader acceptance.

[1]Contact information:  Scott T. Meissner, volunteer Adjunct Professor, University of the Philippines Cebu, Department of Biology and Environmental Science, College of Science, Gorordo Avenue, Lahug, Cebu City  6000, The Republic of the Philippines.
e-mail:  stm4@cornell.edu
    Contribution:  The author is the sole contributor to this article.
    Competing financial interest:  The author declares no competing financial interest.

**1.  Introduction.**
**2.  Lipid pores suggested to be induced by proteins and very common.**
**3.  The Soliton Wave model for the propagation of action potentials.**
**4.  Considering the non-electrical phenomena associated with action potentials.**
    **4a.  Birefringence changes during an action potential.**
    **4b.  The pattern of heat emission and reabsorption during an action potential.**
    **4c.  Change in cell diameter with action potential passage.**
**5.  Are lipid phase transitions common and adaptive?**
**6.  In conclusion.**

**1.  Introduction.**

Over the past two decades there have been suggestions that the Hodgkin-Huxley model [Hodgkin *et al.*, 1952a;  Huxley, 2002] of the electrical features of action potentials should be replaced with other models because it does not take into account several reported phenomena associated with action potentials:  Specifically, the birefringence change in the membrane, the heat emission and reabsorption, and the change in diameter of the cell.  As part of this challenge, one alternative model has adopted a rather new perspective, one which puts more stress on the traits of the membrane lipids rather than on the membrane proteins.  For an overview of some aspects of this proposal see the following:  Andersen *et al.*, [2009], Appali *et al.*, [2012], Heimburg *et al.*, [2005], Heimburg, [2007, 2009], and Lautrup *et al.*, [2011].  These workers have also suggested that lipid pores might account for many of the ionic currents currently associated with protein-based ion channels [Blicher *et al.*, 2013;  Heimburg, 2018].  Both of these proposals are interesting, creative, and potentially paradigm shifting.  But in order for these proposals to gain broad acceptance, then clearly there is a need to test these proposals, in living cells.

Therefore, the main purpose of this review is to place these proposals into a context of the existing biological literature, identify some of the interesting assumptions and predictions of these proposals, and also to suggest some ways in which specific aspects of these new proposals might be tested.  Also, there will be a consideration of the three non-electric phenomena associated with action potentials.  Alternative explanations for these phenomena, largely dependent on features of the membrane proteins and perhaps consistent with the Hodgkin-Huxley model, will be raised.  Then an essential assumption of the soliton wave model for action potentials, that lipid phase transitions are common and adaptive, will be examined.  Whether these proposals achieve broad acceptance, or not, in the biological community will largely depend on the results of a broad variety of additional tests and observations.

**2.  Lipid pores suggested to be induced by proteins and very common.**

Some studies of artificial phospholipid bilayers, and their transition from a liquid-crystalline phase to a gel phase, have noted that lipid pores often appear during such phase transitions.  Such pores are noted to open and close on time scales similar to that seen for some ion channels [Appali *et al.*, 2012;  Blihert *et al.*, 2013;  Gallaher *et al.*, 2010;  Gutsmann *et al.*, 2015;  Heimburg, 2010, 2018;  Laub *et al.*, 2012;  Mosgaard *et al.*, 2013a, 2013b, 2015;  Wodzinska *et al.*, 2009;  Wunderlich *et al.*, 2009].  These findings are not novel as the formation of pores, and so a rise in conductance, during lipid bilayer phase transitions has been known for decades [Antonov *et al.*, 1980;  Chakrabarti *et al.*, 1992;  Deamer *et al.*, 1986;  Feigin *et al.*, 1995;  Nagle *et al.*, 1978;  Papahadjpoulos *et al.*, 1973;  Wu *et al.*, 1973].  Various influences, other than a phase transition, have also long been known to induce pores in lipid membranes; including electrical fields leading to electroporation [Meissner, 1998;  Velikonja *et al.*, 2016;  Wegner, 2013], or the presence of certain peptides [Almeida *et al.*, 2016;  Bernheimer *et al.*, 1986;  García-Sáez *et al.*, 2007].  Thus lipid pores can indeed form in lipid bilayers, and in biological membranes.

The suggestion has also been made that since pores and channels can show similar conductances under voltage clamp conditions, it may be impossible to tell these two apart [Gutsmann *et al.*, 2015].  This point is well taken, as any ion channel reconstituted into an artificial lipid bilayer, which is then taken into a phase transition state, might well show ion conductances higher than would be expected from just the channel alone.  This higher conductance would likely be due to the phase transition-induced lipid pores and their addition to the observed conductance.  Such a situation might account for the rise in conductance reported in a study of a $K^+$ ion channel reconstituted into artificial lipid bilayers when the bilayer was in the state of phase transition [Seeger *et al.*, 2010].  Thus, in such cases, it might be incorrect to attribute all the conductance just to the channel itself.  However, there are a number of ways to tell whether the conductance seen is through a lipid pore versus through a protein-based ion channel.  One is that since the pores will most likely form during phase transition of the lipid bilayer, then placing the bilayer under conditions where the phase transition is unlikely to happen should reduce the current observed due to lipid pores.  But any ion channel-based conductance would be expected to occur even in the absence of a lipid phase transition, and such an observation might suggest that lipid pores are not accounting for the conductance seen in such a situation.  Also the pores formed during lipid phase transition may not show the same ion selectivity as is often associated with many ion channels [Hille, 1984].  Thus the similarity between lipid pores and channels is an interesting superficial feature, but it can be worked around in experimental studies by these and a number of other ways.

One new proposal is the suggestion that the proteins which comprise ion channels might themselves induce the formation of lipid pores in the membrane [Heimburg, 2010;  Gutsmann *et al.*, 2015;  Mosgaard *et al.*, 2013b].  This is not so far-fetched a suggestion as it may at first seem.  There are various antibacterial peptides which are known to induce the formation of non-selective protein-lipid hybrid pores [Ros *et al.*, 2015;  Tournois *et al.*, 1991].  It is also known that some membrane proteins have conserved lipid-binding domains [Barrantes, 2015].  Thus there is nothing new in the suggestion that proteins might induce pore formation.  However, the advocates for this proposal argue that when, for instance, a sodium ion channel is reconstituted into a phospholipid bilayer the current observed would be passed through a sodium ion specific lipid pore which this protein induces in the lipid phase, rather than through the center of the protein itself.  This view implies that the ionic currents are passed through lipid pores, whose formation may be regulated to some extent by the proteins we currently regard as ion channels.  Thus any inhibitor of the ion channel's current is argued to act via binding to the protein, altering its conformation, and so changing its ability to induce a lipid pore [Heimburg, 2010, 2018].  This view would argue that the specific ionic conductances of different channel types would, then, be due to the induction of lipid pores with distinct ionic selectivity.  For instance, reconstitute the protein of a K$^+$ channel into a phospholipid bilayer, and that protein would organize the lipids in such a way that the resulting lipid pore would be ion selective and pass mainly K$^+$.

This is obviously a paradigm-altering proposal, and there are ways in which it might be tested.  For instance, there is one critical difference between the present ion channel view, and this new induced pore concept:  In the channel view the central passage through the channel protein passes the ions, determines the ion selectivity, and so alteration of the amino acid residues along that passage deep within the protein should greatly alter the ion conductance or selectivity.  In the lipid pore view, in order to organize the lipids into a lipid pore the surface residues of the protein must interact with the lipids, and so changes to those surface residues would be expected to alter the ion conductance.  Thus a test of this proposal of induced lipid pores is implied.  First, take a well characterized ion channel's gene [Aziz *et al.*, 2002;  Huynh *et al.*, 1997;  Shen *et al.*, 2017a] which can be altered in such a way that mutant versions of proteins are made;  ideally where each mutant has just one residue changed along the lining of its central channel.  One could do various control studies to confirm that the changes made alter only the internal structure and not the external shape of the protein, and so select specific mutations that fit this criterion.  For instance, it might be worth confirming that this alteration to the protein's interior does not affect its influences on properties of the surrounding lipids, such as their fluidity or phase transition tendencies.  Then, under the current protein channel view, this altered protein should not show its normal ion conductance when reconstituted into a bilayer.  But according to the new proposed induced lipid pore view this internal change should not greatly alter the protein's ability to interact at its surface with lipids, and so a lipid pore should still be induced and ion current still observed.  Thus this proposal is testable, and those who advocate for this view of protein induction of ion specific lipid pores should consider testing their hypothesis.  In order to displace the current view that protein-based ion channels are a major conduit for ion currents, and replace it with this induced lipid pore model, there will need for this and many other tests to be done.  But if the results from the above proposed test supports this new proposal, that would be one step in that direct.  Until such evidence is produced, the current model that ion currents are passed mainly through protein-based ion channels will continue to dominate, rightly so based on the known evidence for it [Hille, 1984].

## 3.  The Soliton Wave model for the propagation of action potentials.

A second area in which some radical new thinking is brought forward is the proposal that an action potential is not as the Hodgkin-Huxley model [Hodgkin *et al.*, 1952a] describes.  Instead a soliton wave, involving the passage of a membrane lipid phase transition down the axon of neurons, is suggested to be the basis for the phenomenon we call an action potential [Heimburg, 2005].  (For some background on solitons, and some applications suggested by others, see the works of Kippenberg *et al.*, [2018], Lomdahl [1984], Lomdahl *et al.*, [1984] and of Layne [1984].)  The change in state of the lipid phase of the membrane with the passage of this solition wave would, it is argued, alter the local thickness of the membrane, and the local surface charge density.  This is suggested to induce a change in the net transmembrane potential in a pattern similar to that of an action potential [Heimburg *et al.*, 2005, 2006;  Lautrup *et al.*, 2011].  One interesting aspect of this model is that it does not require the involvement of any ion channels at all, as it claims that no transmembrane current flow needs to occur.  Thus, no specific ion channel activities would be needed during an action potential based on this soliton wave model [Appali *et al.*, 2012;  Gonzalez-Perez *et al.*, 2014;  Heimburg, 2010;  Vargas *et al.*, 2011].  Thus the soliton wave model for the action potential is a remarkable proposal in many ways, and highly distinctive from the Hodgkin-Huxley view of this phenomenon (Table 1).

**Table 1.**  A comparison of some features of the action potential as envisioned by the Hodgkin-Huxley model versus that envisioned by the soliton wave model[1].

|  | According to the Hodgkin-Huxley model. | According to the Soliton wave model. |
|---|---|---|
| Involves ion channels? | Yes | No |
| Result of action potential collision? | Annihilation | Passage unaltered |
| Uses lipid phase transition? | No | Yes |
| Requires transmembrane ion currents? | Yes | No |
| Accounts for birefringence shift, heat emission and absorption, and cell swelling? | No | Yes |

[1] For fuller details about these and additional comparisons see Appali *et al.* [2012].

Another interesting prediction which the soliton model makes is that two colliding action potentials in a cell should not block each other's passage.  Instead the two oppositely moving action potentials are predicted to pass through each other and continue on [Appali *et al.*, 2012;  Gonzalez-Perez *et al.*, 2016;  Lautrup *et al.*, 2011;  Vargas *et al.*, 2011].  This is rather different from the current view of action potential collision in which annihilation of each of the action potentials would occur (Table 1).  Such annihilation is attributed to the trailing refractory period after each action potential which would block the passage of the oppositely moving action potential for a time.  Such annihilation of action potentials has long been used to test for whether a single neuron was being monitored [Iggo, 1958].  In contrast, concerning action potential annihilation the advocates of the soliton wave model claim that "... compelling evidence for this is not easily found in the literature" [Lautrap *et al.*, 2011], even though, as has been noted [Berg *et al.*, 2017], there are many articles which use action potential annihilation as a means to confirm the recording of a single neuron which may extend to distant locations in the central nervous system [Ferraina *et al.*, 2002;  Kelly *et al.*, 2001;  Movshon *et al.*, 1996], and so identify neuronal connections.  However, several studies done by the advocates of the soliton wave model, present data they suggest illustrate the passage of action potentials past each other, without annihilation, which would be consistent with their soliton wave model [Gonzalez-Perez *et al.*, 2014;  Gonzalez-Perez *et al.*, 2016;  Wang *et al.*, 2017].  However, one criticism of this work is that, in some cases rather than using a single neuron, there was use of a nerve fiber made up of many neurons.  In such a situation there may be some doubt as to whether the two oppositely moving action potentials observed to pass each other were in the same neuron, or instead were in different neurons and so never had a chance to annihilate.  In one of their studies of action potential passage, the advocates of the soliton wave model do state:  "In less than 15% of the preparations, we found annihilation of the pulses"  [Gonzalez-Perez *et al.* 2014; pg. 9].  This might be accounted for, if the electrical stimulations given to induce the action potentials affected only one of five or six neurons in the nerve bundle, as then there would only be about a 20% or less chance of a neuron being stimulated at the other end of the same nerve bundle.  In such a case, over 80% of the time the induced action potentials would be expected to be in different neurons, and so pass each other by without interaction.  This possibility was pointed out in a published comment, which suggested several technical aspects which might improve this test, and noted that action potential annihilation has been reported for many decades [Berg *et al.*, 2017].  In response, the advocates of the soliton model made several comments, including that the suggested use of a higher stimulating voltage, so that all the neurons in the nerve bundle would fire action potentials, would perhaps damage the cells and so block action potential passage by artifact, and they also suggest that the numerous reports of action potential annihilation in the literature may need to be reevaluated [Wang *et al.*, 2017].  It might be worth noting that in addition to action potential collision and annihilation within one axon, there has also been use of what are called 'asymmetric' collisions in which one action potential reaches a strong synapse which induces an action potential in a post-synaptic neuron.  This new action potential in this next neuron in the circuit then goes on to collide and undergo annihilation with an induced action potential traveling in the reverse direction in that second neuron.  In this way the location of synapses can be mapped [Yeomans, 1995].  Such a system might also be a way to avoid some of problems suggested to be associated with electrical stimulus-induced damage?  Be that as it may, the suggestion to test the soliton model by examining its prediction of action potential passage through each other on collision is a good one, but obviously tests of it need to be repeated in a clear single-neuron system where complications can be avoided.  Ideally, a single neuron, well isolated and stimulated using intracellular electrodes, also perhaps under whole cell patch clamp, would allow the induction of action potentials using stimuli closer to those seen normally *in vivo*.  Recently, a test of action potential collision was done involving induced action potentials in the giant internode cells of *Chara* sp.  In this study action potential annihilation was observed to occur in all cases [Fillafer *et al.*, 2017], though is should be noted that this cell was not a neuron.  However, so far, the balance of the data on action potential annihilation versus passage seems to favor the established Hodgkin-Huxley model, and to argue against the soliton wave model.  More testing would always be welcome.

Another difference between the soliton wave model and the Hodgkin-Huxley model of action potentials is that the former makes the prediction that no transmembrane currents (Table 1) are needed in the generation or propagation of an action potential [Appali *et al.*, 2012; Gonzalez-Perez *et al.*, 2016;  Heimburg *et al.*, 2006;  Mosgaard *et al.*, 2013b;  Vargas *et al.*, 2011].  This would seem to imply another rather simple test of this soliton model.  An action potential generated according to the Hodgkin-Huxley mechanism would require the presence of at least one type of voltage sensitive sodium ion channel in the axon of the neuron.  There are several subtypes of such channels which have been identified, as well as some inactive mutant versions [Pau *et al.*, 2017;  Penzotti *et al.*, 1998;  Shen *et al.*, 2017a].  Obviously, an action potential in the Hodgkin-Huxley mode would not be expected to be generated if certain voltage gated sodium ion channel activities were blocked in a neuron.  However, if, as suggested by the soliton model, such ion channel activity is not needed, then even in its absence action potentials should be observed.  To completely remove such ion channel activity one might wish to alter neuronal development.  Therefore, if a neuronal stem cell in culture had its genes for voltage sensitive sodium ion channels altered so that they were either not expressed at all, or so that they coded for a mutated version of this protein which either lacked its voltage sensitivity or had its internal ion conducting region blocked, then the needed channel activity would be removed.  This might allow for a discriminatory test between these two models once the cell matured.  If no action potentials are observed in such an altered neuron, then the Hodgkin-Huxley model would be supported.  But if action potentials were generated even in the absence of this ion channel activity, that would be strong evidence in favor of the soliton wave model.  It should be noted that such a test could be set up to leave a non-conductive version of this ion channel still in the membrane, so any interactions it might have with the membrane lipids which might relate to the formation of the soliton wave would presumably still be present.  Such an explicit test seems not to have been done, yet, by the advocates of the soliton wave model, but should be considered.  One close approximation of this test, however, is a study which examined the expression of the voltage sensitive sodium ion channel's gene during the differentiation of neurons grown in culture.  That study found that in the absence of expression of the gene for this channel, no action potentials were observed, but with increases in expression of this ion channel's gene the cell did develop to a point where action potentials could be observed to be generated in the cell [Liu *et al.*, 2012].  This would seem to support the Hodgkin-Huxley model, and challenge the soliton wave model.  Therefore, those who would wish to find support for the soliton wave model might wish to consider doing an explicit test of their model, focusing on whether the expression of this sodium ion channel's gene, or other ion channel gene expressions, are truly needed for action potential generation by a cell or not.

**4.  Considering the non-electrical phenomena associated with action potentials.**

One argument which the advocates of the soliton model make is that the Hodgkin-Huxley model is unable to account for several phenomena (Table 1) reported to be associated with action potentials;  namely the observed changes in birefringence, the pattern of heat emission and reabsorption, and the minor cellular swelling [Appali *et al.*, 2012;  Heimburg *et al.*, 2005, 2006]. The soliton model approaches this issue from the perspective of how properties of the lipids in the biological membrane, especially a liquid crystalline to gel phase transition in the lipid phase of the membrane, might account for these three phenomena.  They make the argument that because the Hodgkin-Huxley model does not account for these phenomena it is some manner incomplete, and so suggest that it needs to be replaced [Andersen *et al.*, 2009;  Gonzalez-Perez *et al.* 2016; Heimburg *et al.*, 2006].

This gets to the question of what makes a scientific model a good and useful model. Clearly the original Hodgkin-Huxley model did not cover these phenomena of birefringence, heat patterns, and cell swelling, though the authors were aware of some features their model did not cover [Hodgkin *et al.*, 1952a].  Rather than attempt to create a 'model of everything' they choose to focus on the electrical elements associated with the action potential as a beginning.  This is a typical approach in science;  first an initial model is created which accounts for a selected set of observations, and then, if it is a 'useful' model, new additions can be made to it.  Meunier *et al.* [2002] describes some additions which have been made to the original Hodgkin-Huxley model, and suggests that it has indeed been useful over time.  Thus, being 'incomplete' is no reason to reject a model, unless the model is found to be in clear contradiction with solid observations. Indeed, the original model of genetics put forward by Mendel, and the concept of evolution proposed by Darwin, and even the elements of physics presented by Newton can each be argued to have been 'incomplete' in some ways.  But even though 'incomplete' these models have served us well as organizing templates from which more expanded models have been devised in response to the demands of newly observed phenomena.  From this perspective, few scientific models can truly be considered to be 'complete' in that they seldom account for all associated phenomena.

In point of fact, it could be argued that another example of an 'incomplete' model is the soliton wave model for action potentials itself.  Its proponents have chosen to focus on only certain features associated with the action potential, but not all of them.  As its own proponents note, it does not, yet, account for any transmembrane ion currents [Andersen *et al.*, 2009], yet there are specific ion currents which are well documented [Fleidervish *et al.*, 2010;  Hodgkin *et al.*, 1952b, Hodgkin *et al.*, 1955] to occur during an action potential.  Nor does the soliton wave model account for the phenomenon of saltatory conduction which has been shown to occur during some action potential propagations [Hodgkin, 1937].  Nor does it account for how action potentials are able to be observed across a wide range of temperatures in some invertebrates [Hyun *et al.*, 2012].  Neither does it account for why action potentials could be observed by Howarth *et al.* [1968] in a rabbit vagus nerve held at 5°C;  a temperature so far below what is normal for a rabbit's body temperature that one would assume a soliton wave process set up to use a lipid phase transition at normal body temperatures would become rather unlikely?  Those are just a few of the features of action potentials which are not yet accounted for by the soliton wave model.  However, this lack on the part of the soliton wave model is no more a reason to reject it, than are the limitations of the original Hodgkin-Huxley model a reason for rejecting that model either.  Rather what needs to be considered is whether each model can be modified, expanded, altered, or otherwise adapted, and so shown to be consistent with phenomena which fall outside of their original borders.  That would show their 'usefulness.'

Therefore, in this next section, each of these three phenomena associated with action potentials (birefringence changes, heat, and cell swelling) will be considered.  The good people who advocate for the soliton model make arguments that these phenomena can be accounted for by a lipid phase transition [Heimburg, 2005].  But since biological membranes have a significant protein content, whether these phenomena might also be accounted for through changes in the proteins will be considered.  In doing this some possibilities will be raised which, if verified, might operate in a manner largely consistent with the Hodgkin-Huxley model.  It is hoped that this will give the proponents of the soliton model a set of alternate hypotheses to consider, rejection of which they might wish to demonstrate as one means of gaining some support for their argument that lipid phase transitions are involved.  In contrast, advocates of the Hodgkin-Huxley model might also consider these hypotheses as possible means to account for these phenomena in a manner consistent with the Hodgkin-Huxley model.  Thus the purpose here is to suggest hypotheses which others might wish to consider for further investigation.

### 4a.  Birefringence changes during an action potential.

There have been reports that with the presence of an action potential there is a birefringence shift in, or near, the membrane of the cell [Cohen *et al.*, 1970;  Tasaki *et al.*, 1968; Tasaki, 1999a].  Any shifts in the molecular ordering of the membrane lipids during an action potential might contribute to shifts in the birefringence seen during an action potential.  It is well known that phospholipid bilayers, under certain conditions, can show a rise in birefringence due to the ordering of molecular orientations in the gel phase when transitioning from the liquid crystalline phase [Mishima *et al.*, 1996].  Thus the soliton model assumes that there is a liquid crystalline to gel phase transition happening during an action potential, with the gel phase having a much higher ordering of the membrane lipids [Appali *et al.*, 2012;  Heimburg *et al.*, 2005] and this implies that a transient increase in birefringence during the action potential might be expected.

In contrast, some reports note that the birefringence shift seen during an action potential is often tightly associated with changes in the membrane potential which occur during the action potential [Cohen *et al.*, 1969;  Cohen *et al.*, 1974;  Foust *et al.*, 2007;  Landowne *et al.*, 1983], and that there may be involvement of conformational shifts in the membrane's macromolecules [Kobatake *et al.*, 1971].  One study [Cohen *et al.*, 1971], after confirming the birefringence shift seen during an action potential, found that a 50 mV hyperpolarization from the neuron's resting potential could induce a birefringence shift without initiating an action potential, suggesting that this birefringence shift may be related to the membrane potential changes alone.  There have also been attempts to relate protein structure to the birefringence of the macromolecule so that shifts induced via applied electrical fields (i.e. electric birefringence) can be estimated [Pantic-Tanner *et al.*, 1999].  Taken together, these reports seem to suggest that it is the electric field change which may be able to induce a birefringence shift, even without either an action potential or a lipid phase transition being present in the membrane.

**Table 2.**  Non-electrical features associated with an action potential.  Possible ways to account for each in a manner consistent with the Hodgkin-Huxley model, compared to how the soliton wave model accounts for each phenomenon.

|  | Possible means consistent with the Hodgkin-Huxley model. | According to the Soliton wave model. |
|---|---|---|
| Change in birefringence: | Ordering of proteins via the Kerr effect. | Altered lipid order with phase transition. |
| Heat emission followed by heat absorption: | Capacitive charging and discharging, membrane protein conformational changes, and possible changes in protein hydration state. | Capacitive charging and discharging, and heat change with lipid liquid-crystalline to gel phase transition. |
| Transient cell swelling: | Osmotic effects. Changes in protein conformation and hydration, or membrane flexing, with shifts in membrane potential. | Change in bilayer thickness with phase transition. |

Based on this close correlation of the birefringence change with changes in the membrane potential it has been suggested that the Kerr effect might be involved [Beams *et al.*, 1927;  Cohen *et al.*, 1969;  Cohen *et al.*, 1971;  Foust *et al.*,  2007].  The Kerr effect [Weinberger, 2008] is where dipole molecules, especially macromolecules, which are able to flex or shift position, will do so in response to a sudden shift in an imposed electric field.  The shifting by the various macromolecules are thus aligned by the electric field, and the net result can be an increase in organization of the macromolecules, producing an electric birefringence [Kooijman *et al.*, 1997; Krause *et al.*, 1980].  Thus in the case of the cell membrane of neurons, the macromolecules most likely to be involved might be the various plasma membrane proteins.  Obviously, membrane proteins have been reported to shift their structure in response to changes in the membrane potential, as seen by voltage sensitive gating domains of many membrane proteins [Aziz *et al.*, 2002;  Cha *et al.*, 1997;  Shen *et al.*, 2017a;  Whicher *et al.*, 2016].  Thus the changing of membrane protein structure with changes in membrane potential during an action potential is perhaps to be expected, and the birefringence shifts seen would then reflect such altered protein conformational states.

Thus to account for the shift in birefringence seen during an action potential, in a manner consistent with the Hodgkin-Huxley model, the hypothesis might be that the membrane proteins are aligned, through the Kerr effect, in response to the membrane potential shifts which occur during the action potential (Table 2).  Such a hypothesis is testable in various ways, including the comparison of birefringence changes of artificial lipid membranes, with and without membrane proteins reconstituted into them, upon exposure to imposed membrane potentials in such a way that no lipid phase transition occurs.

### 4b.  The pattern of heat emission and reabsorption during an action potential.

Another phenomenon associated with action potentials involves the heat changes which occur.  There are, generally, two ways to estimate the heat released with an action potential.  The first, which uses the Hodgkin-Huxley model as a starting point, is to calculate what the net heat release should be based on the sum of the expected ion flows associated with an action potential.  This has been done by Margineanu $et$ $al.$ [1977] producing an estimated net heat of 49 nJ/cm$^2$ of membrane per action potential, while another estimate [Yi $et$ $al.$, 2016] suggests values from 160-350 nJ/cm$^2$.  Obviously, the values might be expected to vary based on cell types and cell morphology.  As a possible upper estimate, one study [Pissadaki $et$ $al.$, 2013] considered a highly branching human SNC neuron, which had a total axon length of 4.14 m, and suggests that the net energy released would be roughly equivalent to the hydrolysis of 9.36 X 10$^{10}$ ATP molecules per action potential fired in this cell.  The second way to estimate the heat shifts associated with an action potential is to measure it directly.

Two studies [Abbott, 1960;  Howarth $et$ $al.$, 1968] both estimated the net heat from a single action potential to be roughly one μcal/g of axon.  The rather surprising thing about the direct measurements of the heat changes associated with an action potential, however, is that there was first an emission of heat, followed by an absorption of heat [Abbott, 1960;  Howarth $et$ $al.$, 1968].  Generally, the heat emitted seems to be slightly greater than that absorbed.  For instance, Howarth $et$ $al.$ [1968] reports the the emitted heat is in the range of 7.2 μcal/g, while the heat absorbed is roughly 4.9 μcal/g.  It was noted by Tasaki [1999a] that from 50-80% of the emitted heat was eventually reabsorbed, leaving a net heat emission per action potential firing.  Obviously, calculations of net heat production based on theory alone could not predict this pattern of emission and absorption, and this illustrates that there is a need for close observation of phenomena in living material rather than depending merely on theoretical expectations.

The soliton wave model of the action potential attributes this pattern of heat to a liquid crystalline to gel phase transition, and its reversal, by the lipids in the axon membrane.  Studies done on various phospholipid bilayer model systems show heats of lipid "freezing" and "melting" during such phase transitions [Andersen $et$ $al.$, 2009;  Heimburg $et$ $al.$, 2005;  Heimburg, 2012;  Koynova $et$ $al.$, 2013;  Mosgaard $et$ $al.$, 2013a].  The energy change with this phase transition is reported to be in the range of 20-40 kJ/mol, depending on the phospholipid system used [Andersen *et al.*, 2009;  Koynova *et al.*, 2013;  Heimburg *et al.*, 2005;  Heimburg, 2012].  When such a phase change occurs, it would be likely to alter the width of the membrane, and so it is argued that with a lipid phase change there might also be a change in the membrane capacitance [Andersen *et al.*, 2009].  Estimates of heat from changes in membrane capacitance-induced shifts in capacitive charging and discharging have been made [Abbott, 1960;  Howarth *et al.*, 1975;  Margineanu *et al.*, 1977], making this also a reversible process occurring during the passage of an action potential.  Thus the heat emitted by an action potential would be largely reabsorbed with its passage, and the advocates of the soliton wave model therefore suggest that action potentials may occur without any net heat release - as an adiabatic process [Andersen *et al.*, 2009;  Appali *et al.*, 2011;  Gonzalez-Perez *et al.*, 2016;  Heimburg *et al.*, 2006;  Heimburg 2009;  Mosgaard *et al.*, 2013a;  Vargas *et al.*, 2011].  In this way the soliton model is said to account for this pattern of heat emission and reabsorption, and does so, they claim, in a manner which the Hodgkin-Huxley model cannot account for [Andersen *et al.*, 2009;  Appali *et al.*, 2012;  Heimburg *et al.*, 2006;  Gonzalez-Perez *et al.*, 2016], and this is one of the reasons, they argue, that the soliton wave model should be considered.

The question of whether or not the Hodgkin-Huxley model could account for the heat emission and reabsorption seen during an action potential has been considered by several studies. All of them conclude that the flow of ions down their electrochemical gradient during the action potential does give off heat, but that this dissipative heat is not enough to account for the large initial heat emission reported at the start of the action potential [Abbott, 1960;  Andersen *et al.*, 2009;  Howarth *et al.*, 1968;  Howarth *et al.*, 1975;  Margineanu *et al.*, 1977].  So in addition to the heat from ion flows, and any small contribution of heat changes from capacitative charging and discharging, it has been suggested [Margineanu *et al.*, 1977] that reversible processes involving the membrane proteins might be a good candidate to examine as a source of both heat emission and reabsorption.  This could involve protein unfolding which would be reversible, and induced by the shifting membrane potential associated with the action potential.

It has long been known that electric fields can induce conformational shifts in intrinsic membrane proteins.  For instance, Stevens [1978] estimated that a 100 mV potential change across a 10 nm wide membrane might induce enough pull on an electric charge in a protein to shift it by 2.5 nm.  In their calculations Margineanu *et al.* [1977] assumed a reversible heat of unfolding in the range of 10 kcal/mole of protein to be involved.  It might be that the domains of the proteins which extend out of the membrane might be particularly sensitive to unfolding, as there is a report that extramembrane domains may experience enthalpy changes of up to 74 kcal/mol of protein with changes in folding [Powl *et al.*, 2012].  A study of the unfolding of bacteriorhodopsin [Curnow *et al.*, 2007] suggests that the unfolding of each of the membrane spanning helical regions might have an enthalpy change of 20 kcal/mol.  Sanders *et al.* [2018] suggests that a range of free energy of the partial unfolding of the α-helical sections of a membrane protein might range from 3-95 kcal/mole.  For globular soluble proteins, estimates of the energy of unfolding are also in a similar range [Bakk *et al.*, 2001;  Saini *et al.*, 2010].  A study of the conformational changes of the MsbA transport protein going through its transport cycle found that energy changes of hundreds of kcal/mole were evident [Moradi *et al.*, 2013], which certainly suggests that there may be a significant amount of energy shift involved in protein conformational changes.  In addition, if when a protein takes on a new configuration it alters the extent to which it interacts with water, then there might also be contributions from shifts in the state of hydration of the protein regions.  Some recent estimates of the solvation free energies of individual amino acids suggests that having proper values for such hydration energies may greatly influence the resulting estimates of folding of proteins upon hydration [König *et al.*, 2013].

Given the above reports of energy changes with shifts in protein conformation, an argument for how the pattern of heat emission and then absorption might be accounted for in a manner consistent with the Hodgkin-Huxley model might go roughly as follows:  With the reversal in membrane electrical polarity associated with the action potential, any membrane protein with a high dipole state, and some flexibility, would likely have its conformation shifted.  This shifting of the protein charges would include capacitative charge shifts, and by moving to a new lower energy state might result in a release of heat.  With the passage of the action potential these membrane proteins would shift back to the original conformation under the resting potential of the membrane, perhaps absorbing energy in the process.  There might be water molecules exchanged onto, and then off of, the reconfigured proteins during this process as well. And, of course, heat may be emitted by the flow of ions down their electrochemical gradient, though this would only be seen as heat emission.  Obviously, this argument is currently mainly conjecture, and is in need of further study.  But if the emitted heat is due to shifts in protein conformation and/or protein hydration, which in turn are related to changes in the membrane potential, then anything which changes the membrane potential might influence the extent of heat emission.  This is indeed what was observed in one study when the external concentration of sodium ions was lowered, which would be expected to lower the voltage swing of the action potential generated by the neuron, and was noted to result in lower levels of heat emission [Howarth *et al.*, 1968].

Thus it may be possible that the pattern of heat emission and reabsorption reported to be associated with the action potential might be fully consistent with the Hodgkin-Huxley model (Table 2).  If this is confirmed, then it might be a small and appropriate step to expand that model to include this phenomenon within its framework.  Clearly, the advocates of the soliton wave model need to not only show that their model accounts for the observations well, but that it also does so better than reasonable alternatives.  The above suggestion is one such alternative, and as of yet it has not been addressed in depth by the advocates of the soliton wave model who seem to favor looking to just lipid phase changes in biological membranes and to ignore possible roles of the membrane proteins.  If the advocates of the soliton wave model are able to disprove the above hypothesis, they will perhaps strengthen the case for their model to some extent.  Those who advocate the Hodgkin-Huxley model might also wish to explore this hypothesis, to see if it may be confirmed to be consistent with their model.

### 4c.  Change in cell diameter with action potential passage.

The third observation often associated with an action potential is that of the transient swelling of the cell observed during an action potential [Tasaki, 1999a, 1999b].  The extent of the cell swelling differs with reports.  Hill *et al.* [1977] in a crayfish giant axon observed a swelling in the range of 3-25 Å with action potential passage.  Another study, using a nerve from a crab, [Iwasa *et al.*, 1980] noted a swelling in the range of 50-100 Å.  Chéreau *et al.* [2016] observed increases in the axon diameter in the range of 100-300 nm.  While, a study of the action potential in the giant internodal cell of *Chara braunii* [Fillafer *et al.*, 2018] reported a range of swelling in cell diameter associated with the action potential of 1-10 μm.  Fields *et al.* [2010] observed axon swelling of under 0.5 μm with action potential passage, and also reports that this cell swelling induced ATP release out of the cell via volume-activated anion channels.  Thus there seems to be quite a range of variation between taxa in terms of just how much cell swelling is associated with an action potential.  Given the range of the reported cell swellings, some caution is called for in terms of possible causes as several influences might be involved.

  The advocates of the soliton wave model for the action potential suggest that the transient increase in cell size during an action potential is associated with a liquid-crystalline to gel phase transition in the lipids of the excited cell [Heimburg *et al.*, 2005, 2006;  Heimburg, 2012, 2018].  Such a change in thickness can be inferred from the change in capacitance of a phospholipid bilayer after phase transition [Toyama *et al.*, 1991;  White, 1970].  One phospholipid bilayer has been reported to have a thickness of 47.9 Å when in the gel phase, which drops to a thickness of 39.2 Å when in the fluid (liquid-crystalline) phase [Heimburg, 1998].  And changes in the thickness of phospholipid bilayers upon phase transition has been reported in other studies [Koynova *et al.*, 2013;  Mosgaard *et al.*, 2015, Pagano *et al.*, 1973, Stevenson *et al.*, 2017], including one in which atomic force microscopy was used to monitor the phospholipid bilayer thickness change across the phase transition [Goksu *et al.*, 2009].  Obviously in a cell, with an upper and lower cell membrane, this displacement would be expected to be doubled if both the bottom and top cell membranes underwent phase shifts at the same time.  Gonzalez-Perez *et al.* [2016] used atomic force microscopy on a neuron from lobster, and found displacements in the range of 0.2-1.2 nm associated with action potential passage.  One question concerning that work relates to what the atomic force microscopy probe might have been actually touching.  If this probe was overly broad, then it might perhaps be more likely to have touched surface proteins which extend above the plane of the cell membrane rather than the lipids directly?  Whether or not any phase-shift associated change in the thickness of the lipid phase would then be expected to alter the height of such membrane proteins should perhaps be examined.

  In addition to a possible lipid phase transition altering membrane, and so cell, thickness, there are several other plausible means by which the cell diameter might be altered during an action potential.  One possibility may be associated with osmotic effects.  It is noted that cells can swell with ionic fluxes [Sorota, 1992], and that such swelling can alter some aspects of the action potential itself [Decher *et al.*, 2001].  Mosbacher *et al.* [1998] estimates that the ion fluxes out of a voltage clamped HEK cells over the course of twenty seconds could cause the cell to shrink in diameter by 200 nm.  So could the transient ion flows associated with an action potential induce local swelling in this manner?  Another possibility relates to the fact that biological membranes are comprised of large amounts of proteins, and so the possible role of shifts in protein conformation during an action potential would also seem to be relevant.  There are reports that some potassium ion channels are rather flexible [Mosbacher *et al.*, 1998], with Kowal *et al.* [2014] reporting the use of atomic force microscopy to observe that the binding of a ligand to a $K^+$ channel altered the height of that protein by 1.5 nm.  Atomic force microscopy was used in another study on a voltage dependent anion channel isolated from human mitochondria, and found that it was an extremely flexible membrane protein [Ge *et al.*, 2016].  A study using cultured HEK cells, also using atomic force microscopy, found that some ion channels could under go conformational shifts in their vertical height above the membrane in the range of 1.4 nm, and that just the voltage sensing S4 domain alone in such channels often shifted by 20 Å [Beyder *et al.*, 2009].  Thus, could some of the change in cell thickness be due to changes in membrane protein conformation during the time of action potential passage?  Yet another plausible source of change in cell diameter might be that of electroflexion [Petrov, 2006].  Here a shift in the membrane potential can induce changes in membrane curvature, with some sections of the membrane reported to bulge out as much as five nanometers [Mosbacher *et al.*, 1998], and other sections suggested to be anchored to the cortical cytoskeleton [Beyder *et al.*, 2009;  Zhang *et al.*, 2001].  A study of HEK cells under whole cell patch clamp conditions, and which used atomic force microscopy to monitor the electroflexion, found that for each 100 mV of applied hyperpolarization roughly one nanometer of membrane flexing might be induced [Beyder *et al.*, 2009].  Another possibility is raised by El Hady *et al.* [2015], who suggests that the electric field shifts associated with the action potential might affect the underlying cytoskeleton and so alter the tensions applied to the cell membrane, creating a surface swelling.  Finally, there are changes noted to occur in an axon, and at synapses, during the passage of an action potential involving increases in the cell membrane surface area attributed to vesicle fusions [Chéreau *et al.*, 2016], which might result in altered axon surface shape and diameter as well.

Thus there are many possible means by which the cell swelling seen during action potential passage might be produced.  It should be noted that many of the above options could, in principle, operate in a manner consistent with the Hodgkin-Huxley model (Table 2). Of course, it is entirely possible that several of these  mechanisms may, to some extent, operate together to contribute to the observed cell swelling seen during an action potential.  Clearly further work is needed in this area.  Thus, here again is an alternative hypothesis which the good people who advocate for the soliton wave model need to attempt to refute, if they can, so that their proposal that a lipid phase transition accounts for this change in cell diameter can be strengthened.  And those who support the Hodgkin-Huxley model of the action potential may wish to examine some of these phenomena to see if any of them can be confirmed to operate during an action potential.

## 5.  Are lipid phase transitions common and adaptive?

For the action potential to be accounted for by the soliton wave model, the assumption that lipid phase transitions occur regularly in the cell membranes of neurons and of other excitable cells is critical.  Given the importance of this assumption, a brief review of reports of lipid phase transitions from the biological literature would seem to be in order.  It should be noted that there is no doubt that imposed changes in temperature [Verkley *et al.*, 1975], pressure, tension [Imam *et al.*, 2016], ionic strength, water potential, or other added factors [Lenné *et al.*, 2007;  Meerschaert *et al.*, 2015], can influence whether or not biological membranes display lipid phase transition, or the formation of lipid pores [Heimburg, 2007]. For instance, Shen *et al.* [2017b] found that feeding cultured cells highly saturated fatty acids could induce a phase transition in the membrane of the endoplasmic reticulum, where most new phospholipids are assembled in animal cells. However, they also note that long term exposure to such fatty acids ultimately led to cell death. The work of Yu *et al.* [2015] reports that, through electroporation of isolated and cultured neurons, DNA could be delivered via induced lipid pores into some cells for expression in them, with only 40% or less of the electroporated cell dying from the treatment. A study which examined what temperatures best promote preservation of bovine oocytes [Arav *et al.*, 1996] found that cooling below the standard 22°C storage temperature induced a membrane lipid phase transition in the cell membranes leading to damage to these cells, resulting in their rupture and loss. A study of the antibiotic daptomycin [Müller *et al.*, 2016] suggested that its mode of action involved the induction of a lipid phase shift towards lower fluidity in the membrane of the treated bacterial cells; killing them. Together these reports support the general consensus from the biological literature that phase transitions, and the accompanying lipid pores which may form during them, can be very stressful to cells. This view has much additional support (see Table A1 in the appendix). For instance, there are reports that phase transitioning from a liquid-crystalline to a gel lipid phase can lower or eliminate the activity of some membrane proteins [Armond *et al.*, 1979; Boheim *et al.*, 1980; Gennis *et al.*, 1977; Grisham *et al.*, 1973; Koynova *et al.*, 2013; Murata *et al.*, 1975; Silvius *et al.*, 1980; Zakim *et al.*, 1975]. Also, phase transitions in the plasma membrane of cells have been reported to stress and damage cells, mainly by the induction of non-selective pores in the membrane which can lead to the leakage of small items out of the cell [Amir *et al.*, 2008; Balasubramanian *et al.*, 2009; Crowe *et al.*, 1989; Ghetler *et al.*, 2005; Hendricks *et al.*, 1976; Lin *et al.*, 2014, Oldenhof *et al.*, 2010; Ono *et al.*, 1982; Pasternak *et al.*, 1992; Ragoonanan *et al.*, 2008; Senaratna *et al.*, 1984; Sun, 1999], but also can induce higher rates of lipid peroxidation during phase transition producing damaging products in the cell [Scott *et al.*, 1991]. Several studies have concluded that, in order to maintain viability, cells must keep their plasma membrane in a fluid/liquid crystalline state [Chugunov *et al.*, 2014; Kilin *et al.*, 2015; Koynova *et al.*, 2013; Lockshon *et al.*, 2012; Melchior, 1982; Moein-Vaziri *et al.*, 2014; Morein *et al.*, 1996; Overath *et al.*, 1970]. Some articles question whether lipid phase transitions occur in cells under normal conditions, or if they are ever used to achieve any useful physiological outcomes [Chapman *et al.*, 1974; Crowe *et al.*, 1999; Kilin *et al.*, 2015; Lee *et al.*, 2015; Mabrey *et al.*, 1977; Menon, 2018; Palleschi *et al.*, 1996; Pike *et al.*, 1980; Quinn, 1981; Rao *et al.*, 2014; Sevcsik *et al.*, 2015; Zakim *et al.*, 1975]. Conditions which promote lipid phase transitions seem to be linked to some disease states [Alberti *et al.*, 2016; Maulucci *et al.*, 2017; Reddy *et al.*, 2016]. Finally, there is also a report that when under stress some cells produce proteins which act to inhibit lipid phase transitions [Wu *et al.*, 2000], and so help the cell avoid phase transition-associated damage. Taken in total, these studies imply that in order for the cell to maintain its plasma membrane as a controllable permeability barrier, and for the cell to be viable, there seems to be a need to avoid lipid phase transitions when possible.

There are some reports of different states of lipids in membranes, and transitions between them, which are suggested to perhaps be more benign than that of the liquid-crystalline to gel phase transition.  There are two proposed versions of the liquid-crystalline phase, called a liquid-disordered phase (ld) and a liquid-ordered phase (lo).  This relates to the hypothesis of lipid rafts (also called membrane nanodomains) [Dent *et al.*, 2016;  Karnovsky *et al.*, 1982;  Lingwood *et al.*, 2010;  Rao *et al.*, 2014].  A study by Garcia-Manyes *et al.* [2010] using atomic force microscopy reports that different regions of a phospholipid bilayer can have different resistance to 'punch-through' by the probe tip in a manner which may reflect differences in the bilayer between areas in ld, to ones in an lo, state.  Which suggests the existence of distinct nanodomain regions in the bilayer.  Brown *et al.* [1998] suggests that biological membranes are most likely in a disordered liquid crystalline state, but notes the possibility that some sort of shift to a more ordered liquid crystalline state might occur.  Almeida [2011] suggests that the lo and ld  states might coexist, but argues that, since the energy shift between these two states is about an eighth of that seen between an ld and gel phase, there is no significant heat release with phase transition from the ld to the lo states.  It should be noted that there are significant debates over whether or not these different lipid rafts/nanodomains actually occur in the plasma membrane of living cells, with some presenting arguments and observations which challenge some aspects of the lipid raft concept [Leslie, 2011;  Lu *et al.*, 2018;  Sevcsik *et al.*, 2015].  For a good review of some of the technical challenges of studying such small, and fleeting, areas of the membrane see Elson *et al.* [2010].  Thus the entire lipid raft concept is still under active debate and exploration, and how, or if, it might relate to the soliton wave model of the action potential is still unclear.

In contrast, then, to the notion that lipid phase changes are rare and damaging, the proponents of the soliton wave model of action potentials seem to suggest yet another paradigm shift:  That lipid phase changes occur in biological membranes in an adaptive manner, especially in the cell membranes of neurons during action potentials [Heimburg *et al.*, 2006, 2007;  Heimburg, 2018].  Given the central importance of the assumption of lipid phase changes for their soliton model [Andersen *et al.*, 2009;  Appali *et al.*, 2012;  Heimburg *et al.*, 2005;  Vargas *et al.*, 2011] one would expect that this group would supply supporting evidence, perhaps by citing reports of liquid crystalline to gel phase transitions in neuron plasma membranes - but, oddly, no direct evidence is presented.  In some cases, when they do cite a source [Blicher *et al.*, 2013;  Gallaher *et al.*, 2010;  Gonzalez-Perez *et al.*, 2014;  Heimburg, 2010, 2018;  Lautrup *et al.*, 2011;  Mosgaard *et al.*, 2015;  Wodzinska *et al.*, 2009], what is cited is often an earlier publication from their own group, in which this claim was previously implied.  In some cases they do cite studies from outside of their own research group (as done for instance by Heimburg [2007], Laub *et al.* [2012], and Mosgaard *et al.* [2013b]).  But in those instances, the items cited often relate either to phase transitions imposed on the membrane of *E. coli*, to how phase transitions can be seen in extracted lipids from lung surfactant (which Nag *et al.* [1998] notes is a lipid monolayer), or to other items.  However, they do not cite direct reports of observed liquid crystalline to gel phase transitions in the membranes of neurons themselves under physiological conditions.  Thus no direct evidence is cited in support of the claim that in the plasma membrane of neurons such a phase transition commonly occurs.  Of course, phase transitions can be artificially imposed on a neuron, for instance by cooling it or by altering the lipid composition of its membrane.  But obviously such influences are not used *in vivo* in our nervous tissue, and so the physiological relevance of such imposed transitions is left unclear.  Obviously, it is not enough that lipid phase transitions be shown to be inducible in such cells, if the soliton model is to apply to action potentials in neurons and other excitable cells, then such phase transitions must be shown to actually occur in those specific cells under normal conditions.  In addition, if such liquid crystalline to gel phase transitions are actually shown to occur, then the next question would be how does the neuron survive having items leak out through lipid pores which presumably would be induced with each phase transition?

There are reports of studies of the state of lipids in neurons.  For instance, the studies of Sonnino *et al.* [2015], Bonaventura *et al.* [2013], and Saxena *et al.* [2015] examined the cell membranes of neurons, but none of these studies mention any evidence of a liquid-crystalline to gel phase transition, which are said to be needed for there to be soliton waves in a neuron [Heimburg, 2005].  Thus, it would seem to be essential for the proponents of the soliton wave model to produce positive evidence of this specific phase transition in the membrane of neurons during action potential propagation.  Without such evidence, one of the keystone assumptions of their model, indeed the soliton model of action potentials itself, is left in major doubt.

## 6.  In conclusion.

The good people who advocate for the soliton wave model of the action potential, and for the inducible lipid pore model, are proposing hypotheses which would call for the overturning of several existing paradigms:  1)  That the current flows attributed to ion channels might instead be passing through lipid pores induced by the proteins currently thought to be the ion channels.  2) That the Hodgkin-Huxley view of the action potential should be replaced with their soliton wave model.  3)  Associated with that, they challenge the existence of action potential annihilation upon collision.  4)  And they suggest that lipid phase transitions, rather than being harmful to cells, are used in an adaptive and significant manner during the action potential, and perhaps in other contexts.  To call these suggestions bold would be a vast understatement.  On one level, perhaps, these workers are to be praised for the hard work they have done to raise alternatives to existing dogmas, however, they now need to show significant evidence that their suggested phenomena occur inside of real cells in a manner consistent with their proposals.  Clearly, such broad claims require massive amounts of evidence in order to be convincing.

Part of the reason, perhaps, for their boldness is the 'thermodynamic approach' which they take in producing their models [Heimburg *et al.*, 2006;  Heimburg, 2009;  Lautrup *et al.*, 2011; Vargas *et al.*, 2011].  They state that such an approach allows them to produce "... correct predictions..."  [Andersen *et al.*, 2009; pg. 107].  This might be somewhat of a valid claim, to the extent that their resulting models are more likely to be consistent with known physical/chemical rules, and so help to avoid proposals which might violate such rules.  But, it should be noted, that while it is necessary that a proposed model be consistent with such rules, that alone is not sufficient to demonstrate that life makes adaptive use of aspects of that model's phenomena. After all, there are many phenomena which are fully consistent with physical/chemical rules but which are actively avoided by most forms of life.  For instance, consider the case of liquid water in the cytosol of cells.  With a sufficient drop in temperature, this water can be induced to transition into a solid phase, which can be described via thermodynamics [Khvorostyanov *et al.*, 2004].  But the formation of ice crystals inside a cell can damage cellular structures resulting in the leaking out of cell contents [Barbier *et al.*, 1982], and so to avoid this damage many species produce various internal changes, some at significant expense, in an attempt to inhibit or limit the formation of such ice crystals [Moellering *et al.*, 2010;  Sun *et al.*, 2014;  Wu *et al.*, 2000].  Thus, just because ice formation is possible according to physical/chemical rules, that does not mean that it is used by cells in an adaptive manner.  The same may be said for the phase transitions which many membrane lipids are capable of undergoing.  By thermodynamics such phase transitions may indeed be shown to be possible, but that does not exclude other phenomena also being possible.  And given that the mechanisms underlying the Hodgkin-Huxley model are also consistent with known physical/chemical rules, clearly the mechanisms of the Hodgkin-Huxley model are alternatives which cells have the option to use, rather than making use of solitons, in producing their action potentials.  Therefore, the thermodynamic approach taken by the advocates of the soliton wave model has its benefits, but it does not remove the need to test and confirm the actual occurrence of the phenomena associated with their models inside of living cells.

This review has presented several ways in which aspects of the soliton model of the action potential, and the accompanying proposal of induced lipid pore use in life, might be tested.  Many other tests of these proposals are possible.  In addition, some suggestions have been raised relative to how the three non-electrical phenomena associated with the action potential (birefringence shifts, patterns of heat emission and absorption, and changes in cell diameter) might be accounted for in a manner largely consistent with the Hodgkin-Huxley model.  Clearly more study of these three phenomena associated with action potentials needs to be done.  It is hoped that this review will help stimulate such work.  It will be the results of such tests which will ultimately determine which aspects, if any, of the soliton wave model, and of the proposal concerning lipid pores, become incorporated into our views of life.  Meanwhile, we should acknowledge the value of work which challenges dogma.  And so, even if none of their proposals are found to be upheld, the advocates of the soliton model have provided a useful service in providing some original thinking, and some creative alternative hypotheses relative to existing dogma.  Hopefully they will continue to do so.

**Acknowledgements:**

The author would like to thank the faculty and staff at UP Cebu for their hospitality and for permitting the use of their facilities while writing this article.  Also, thanks is due to Prof. Heimburg for his patient replies to many questions sent his way.

**Appendix:**

**Table A1**.  Quotations (and citing of sources) from studies examining phase transitions in living cells, and their effects. A.)  Phase transition effects on individual protein activities.  B.)  Phase transitions found to induce stress on cells, including leakage of small items.  C.)  The fluid [liquid-crystalline] state is said to be needed for cell viability.  D.)  Phase transitions said to not normally occur in certain biomembranes.  E.)  Phase transitions found to be associated with some disease or illness states.

| A.) | Phase transitions effects on individual protein activities: | |
|---|---|---|
| | "Experiments carried out on membranes from 1:1 (wt/wt) mixture of dipalmitoylglycerol and distearoylglycerol in *n*–decane led to the interpretation that ion carriers become frozen and thus immobile within the membrane phase (6)." | Boheim *et al.* [1980], pg. 3403 |
| | "In our situation the carrier-mediated ion transport is blocked completely below 26-27$^\circ$C, which matches closely the phase $t_c$ of the pure lipid.  Most probably the valinomycin molecules have been frozen out." | Boheim *et al.* [1980], pg. 3406 |
| | "... in general it is accepted that the physical state of the lipid is critical for the activity of many integral membrane enzymes... $C_{55}$-isoprenoid alcohol phosphokinase (232) and $Ca^{2+}$-ATPase from sarcoplasmic reticulum (162) have been shown to require a fluid bilayer to function....  Membrane viscosity is of such importance that organisms, from mammals down to bacteria, have mechanisms for altering the membrane lipid content to maintain a fluid membrane when growth conditions (i.e. temperature) are changed (4, 233)." | Gennis *et al.* [1977], pg. 216 |
| | "The temperature studies suggest that the ATPase must be in a 'fluid-like' environment to function." | Grisham *et al.* [1973], pg. 2635 |
| | "... further supporting the suggestion that the membrane lipids must be fluid for the ATPase to function." | Grisham *et al.* [1973], pg. 2635 |
| | "The transition from liquid crystalline to gel phase, which results in a marked change in the physical properties of lipid bilayer;  also strongly affects the activities of membrane proteins.  Membrane proteins, for example, the $Ca^{2+}$-ATPase, show low activity in gel phase bilayers, due to the effect of gel phase on protein conformation[55]." | Koynova *et al.* [2013], pg. 6 |

| | |
|---|---|
| "At temperatures near 0 C the lipid area of the membrane in the solid state are probably large and this may influence the binding between the phycobilisomes and the thylakoid membranes.  In relation to this it is interesting to note that *Anacystis* is an organism that is especially chilling sensitive.  When the cells are stored at 4 C for 1 hr, they lose the activities of photosynthesis and the Hill reaction (6)." | Murata *et al.* [1975], pg. 796 |
| "Our results indicate that the enzyme is active only in association with liquid-crystalline lipids, exhibiting a significant heat capacity of activation, $\Delta C_p^{\ddagger}$, for the ATP hydrolytic reaction in this case." | Silvius *et al.* [1980], pg. 1255 |
| "The apparent inactivation of the ATPase when its boundary lipids enter a low-temperature, presumably gel-like, state suggests that the overall enzyme-catalyzed reaction involves at least one step that requires a transient displacement or deformation of the lipid molecules around the enzyme, because such as step would be hindered by the more rigid gel-state lipids." | Silvius *et al.* [1980], pg. 1258 |
| "The regulatory properties of the form of UDP-glucuronyltransferase produced by interaction with a rigid lipid environment (the structure of the lipid phase below 16°) preclude its efficient function as a detoxification system *in vivo*." | Zakim *et al.* [1975], pg. 343 |
| "Thus, the function of this membrane-bound enzyme depends in a critical manner on the careful regulation of the bulk phase properties of its lipid environment." | Zakim *et al.* [1975], pg. 343 |
| B.) | Phase transitions induce stress on cells, including leakage of small items: |
| "Chilling susceptibility has been shown to be correlated with the membrane's saturated-to-unsaturated fatty-acid ratio (Arav *et al.*, 2000) and to be associated with the lipid-phase transition (LPT), as described for both sperm (Drobins *et al.*, 1993) and oocytes (Arav *et al.*, 1996)...  [this]...  lipid-phase separation...  ...interferes with membrane function and leads to ion leakage and cell death." | Amir *et al.* [2008], pg. 150 |
| "The reorganization of lipids and changes in its residual membrane conformational disorder due to phase transitions from liquid crystalline to gel phase [4,11,12] or from liquid crystalline to hexagonal phase [3] under different thermal and hydration conditions have been implicated in injury.  The consequences of these phase changes are thought to include increased membrane permeability and lateral phase separation of membrane components." | Balasubramanian *et al.* [2009], pg. 945 |
| "It has been established that for lipid bilayers in liposomal systems, maximal leakage of intra-liposomal contents coincides closely with the phase transition temperature [34].  Hence, it is possible that freezing induced dehydration damage occurs as a result of leakage of cytoplasmic compounds during the phase transition, either during cooling or thawing." | Balasubramanian *et al.* [2009], pg. 950 |
| "Chilling injury occurs when the cell membrane undergoes a transition from the liquid state to the gel state." | Ghetler *et al.* [2005], pg. 3385 |
| "This enhancement [higher leakage of amino acids] is interpreted as evidence for membrane transitions in the plasmalemmas.  The transitions are also considered to be main factors limiting the most effective temperature range for seed germination." | Hendricks *et al.* [1976], pg. 8 |
| "It has been proposed that lipid phase transition (LPT) may have important implications for susceptibility to chilling injury, as LPT can affect cell membrane properties, such as their function and integrity [10,11,12]." | Lin *et al.* [2014], pg. 1 |
| "In addition, lipid phase transitions alter membrane structure and lateral organization [5,19].  Freezing results in both temperature-dependent (thermotropic) and dehydration-induced (lyotropic) membrane phase changes, which are thought to result in lateral phase separation of membrane components and increased membrane permeability for solutes [8,18,20,25]." | Oldenhof *et al.* [2010], pg. 115 |
| "The freezing-induced lyotropic membrane phase transition is different from the thermotropic phase transitions that occur upon cooling and re-warming of cells at suprazero temperatures [8,20]. Both, however, can result in lipid phase separation and rearrangements, which can be detrimental for cells." | Oldenhof *et al.* [2010], pg. 121 |
| "The results of this and previous studies suggest that the chill susceptibility of *A. nidulans* is a result of irreversible leakage of ions from the cytoplasm when the lipids of cytoplasmic membrane are in the phase-separation state at low temperatures." | Ono *et al.* [1982], pg. 125 |
| "The experimental result suggests that the phase-separation state of lipids of the cytoplasmic membrane induces the chilling susceptibility of this alga." | Ono *et al.* [1982], pg. 125 |
| "... we conclude that the chilling susceptibility of *A. nidulans* occurs in a mechanism as follows. First, the primary response which occurs at chilling temperatures is the formation of gel-phase domain in the cytoplasmic membrane, and, in the phase-separation state, the membrane becomes passively permeable to small molecules. Second, the ions and solutes having low mol wt leak from the cytoplasm to the outer medium. Finally, decreases in the concentrations of the ions and solutes in the cytoplasm inactivate the physiological activities of the algal cells. It should be noticed here that the ions and solutes are irreversibly lost from the cells, even if the cytoplasmic membrane regains the state of liquid crystal when the once-chilled cells are rewarmed. This should be the reason why the chilling-induced injury of *A. nidulans* is an irreversible process (11, 12)." | Ono *et al.* [1982], pg. 128 |
| "Thus, gel phase lipid was detected only in those axes which have been previously shown to have reduced viability and increased solute efflux during rehydration (22, 23)." | Senaratna *et al.* [1984], pg. 761 |
| "The formation of gel phase lipid, which is an indication of a lateral phase separation of phospholipids within the plane of the membrane, would thus be expected to contribute to the loss of viability of the axes after dehydration." | Senaratna *et al.* [1984], pgs. 761-762 |
| "Other mechanisms must be involved in causing massive cellular damage in red oak seed tissues during cryopreservation.  One such mechanism might be membrane phase transition due to freezing or freeze-induced dehydration.  Unlike dessication-tolerant seeds, recalcitrant red oak seed were unable to retain a liquid-crystalline phase upon drying and severe dehydration resulted in membrane phase transition (34)." | Sun [1999], pgs. 383-384 |

| C.) | The fluid [liquid-crystalline] state is needed for cell viability: |
|---|---|
| [Notes that keeping the membrane in a liquid crystalline state is] "... vital for each living cell." | Chugunov *et al.* [2014], pg. 1 |
| "... analyses confirm that hydrophobic tail branching modifications render membranes more liquid and less rigid, which is a requirement for the membrane's biological function." | Chugunov *et al.* [2014], pg. 5 |
| "In line with previous reports (31,34,41,54,62,66,83), the Lo-like phase was clearly dominant in live cells, whereas Chol depletion or apoptosis was found to increase the Ld-like phase." | Kilin *et al.* [2015], pg. 2529 |
| "The dominant factor accounting for acquired protection to low temperature damage relates to the membrane lipid composition...  Since increased chain unsaturation strongly reduces the gel-liquid-crystalline phase transition temperature of lipids[13], the increase in the content of unsaturated fatty acids serves as an adaptation mechanism allowing to maintain the membrane in its physiological liquid-crystalline state also at much lower temperatures." | Koynova *et al.* [2013], pg. 6 |
| "Preservation of both the integrity and fluidity of biological membranes is a critical homoeostatic function." | Lockshon *et al.* [2012], pg. 1 |
| "This work provides the first evidence for the existence of a signaling pathway that enables eukaryotic cells to control membrane fluidity, a requirement for division, differentiation and environmental adaptation." | Lockshon *et al.* [2012], pg. 1 |
| "Core components of this sole yeast Rho1/Pkc1/MAPK cascade thus appear to participate in two signaling pathways, one which monitors the integrity of the cell wall and the other which enables membranes to maintain proper fluidity." | Lockshon *et al.* [2012], pg. 7 |
| "Fluidity is a characteristic feature of membranes.  Cellular viability and function is dependent on maintenance of cell membrane fluidity within physiological ranges (Hovath *et al.* 1998, 2008)." | Moein-Vaziri *et al.* [2014], pg. 729 |
| "In one of the commonly employed models for membrane lipid regulation it is asserted that such a regulation is the reason why numerous organisms change the proportion between saturated and unsaturated acyl chains in relation to the surrounding temperature (Hazel and Williams, 1990; McElhaney, 1984). However, it should be recognized that a lipid bilayer being completely in a liquid crystalline state surely has the best prerequisites for maintaining the integrity and the permeability barrier of the membrane and to support full activity of membrane-associated enzymes and transport proteins. It has consistently been shown that living cells prefer to grow with their membrane lipids exclusively, or nearly exclusively, in the liquid crystalline state (McElhaney, 1984). Therefore, *one* aim with the adjustment of the membrane lipid composition may be to avoid the occurrence of gel state lipids in the membrane or at least to reduce the fraction of gel state lipids as much as possible." | Morein *et al.* [1996], pg. 6807 |
| "*E. coli* cells subjected to a significant decrease in the growth temperature therefore have to incorporate a larger fraction of unsaturated acyl chains into its membrane lipids in order to avoid the deleterious effects of having a mixture of lipids in the gel and liquid crystalline states." | Morein *et al.* [1996], pg. 6808 |
| "It is concluded that a liquid-like state of the lipid phase is required for proper membrane function." | Overath *et al.* [1970], pg. 606 |

| D.) | Phase transitions do not normally occur in certain biomembranes. |
| "The presence of cholesterol at sufficient relative concentration has the effect of removing these phase transitions (6, 7).  Those biomembranes which contain large amounts of cholesterol, *e.g.* myelin membranes and also their total lipid extract, do not show these marked endothermic transitions except after some dehydration (8)." | Chapman *et al.* [1974], pg. 2512 |
| "Gramicidin A is an ionophore...  With this molecule, the pretransitional peak is affected at low polypeptide concentrations, suggesting that the packing of the lecithin polar groups has been affected, but in addition to this the heat involved in the main lipid endothermic transition is markedly reduced in a somewhat similar manner to that observed with cholesterol.  It seems reasonable to conclude, therefore, that as with cholesterol the molecule is interdigitated among the lipid chains preventing chain crystallization from occurring." | Chapman *et al.* [1974], pg. 2520 |
| "With mitochondrial membrane the fact (10) that the transition appears to be complete before 37° suggests that at body temperature the lipids of this membrane are completely fluid." | Chapman *et al.* [1974], pg. 2520 |
| "One of the hallmarks of archaeal membrane is the lack of distinct phase transitions and the relatively weak dependence of the membrane's properties on temperature." | Chugunov *et al.* [2014], pg. 4 |
| "Nevertheless, we are confident that it [phase transition] commences during chilling at about 22°C, just below the minimal storage temperature used by blood banks." | Crowe *et al.* [1999], pg. 181 |
| "There is an abundance of evidence that most animal cells contain a large fraction of their membrane lipids as unsaturated phospholipids, with the unsaturated acyl chain in the *sn-2* position.  Such phospholipids are usually thought of as having low phase transitions, often much below 0°C." | Crowe *et al.* [1999], pg. 184 |
| "... the FLIM images of intact and Chol-depleted cells labeled with F2N12S showed a remarkable homogeneity in their pseudo-color distribution (Fig. 7A), indicating that no clear phase separation could be perceived." | Kilin *et al.* [2015], pg. 2526 |
| "We find no evidence of a discrete miscibility phase transition throughout a wide range of temperatures:  14 - 37 °C." | Lee *et al.* [2015] |

Meissner:  Proposed tests of the soliton wave model of action potentials.

| | |
|---|---|
| "Taken together, these data suggest that live cell membrane may avoid the miscibility phase transition inherent to its lipid constituents by actively regulating physical parameters, such as tension, in the membrane." | Lee *et al.* [2015] |
| "Suggesting the cell membrane may be maintained in a different region of the phase diagram and may even actively avoid a temperature-driven miscibility phase transition." | Lee *et al.* [2015] |
| "Despite the fact that lipid mixtures directly extracted from cells exhibit miscibility phase transitions around room temperature,[7-8] no such transition is observed in the live cell." | Lee *et al.* [2015] |
| "There is no evidence for a reversible lipid phase transition at any temperature above $0°$, indicating that the microsomal membrane is in the fluid state under these conditions." | Mabrey et al. [1977], pg. 2929 |
| "There is no evidence for a reversible lipid phase transition at any temperature above $0°$ in the microsomes or extracted lipids....  Because we cannot detect a reversible lipid phase transition in the microsomal preparations we conclude that the membrane is in the fluid state under the conditions studied." | Mabrey *et al.* [1977], pg. 2930 |
| "Sterols ensure that the fluidity of the PM and its barrier function are preserved across a range of environmental conditions [3,9].  They accomplish this by condensing phospholipids with unsaturated acyl chains, and preventing saturated lipids from forming a gel phase." | Menon [2018], pg. 37 |
| "The absence of phase transitions in the same thermal range argues against the hypothesis that the lipid domains previously detected on the sperm surface are produced by lateral phase separation." | Palleschi *et al.* [1996], pg. 197 |
| "The linearity of the temperature dependence of exGP values (Fig. 4) indicates that no phase transitions have occurred within the thermal gradient applied." | Palleschi *et al.* [1996], pg. 200 |
| "... very recent data show that at cholesterol concentrations > 15 mol% coexistence of lipid phase domains cannot be detected in preparations made of mixed phospholipids [28]." | Palleschi *et al.* [1996], pg. 201 |

Meissner:  Proposed tests of the soliton wave model of action potentials.

| | |
|---|---|
| "Winter active ephemerals appear genetically programmed to synthesize a mixture of phospholipids which will not phase separate in the usual growth conditions." | Pike *et al.* [1980], pg. 238 |
| "... it is postulated that it may only be necessary that the phase separation temperature of the lipids be as low or lower than the normal minimum temperatures during the plant's growing season." | Pike *et al.* [1980], pg. 240 |
| "It appears that plants normally growing in a certain environment synthesize a mixture of phospholipids which will not phase separate in the usual thermal regime." | Pike *et al.* [1980], pg. 241 |
| "Even in the case of poikilothermic organisms there is as yet no convincing evidence that lateral phase separations serve any useful physiological purpose." | Quinn [1981], pg. 43 |
| "In artificial membranes the lo phase is insoluble in non-ionic detergents at low temperatures, suggesting that the 'rafts' are detergent resistant.  This has spawned an enormous but misguided literature on this subject [21,22], bringing into question the whole premise of the role of phase segregation in cell membranes [23]." | Rao *et al.* [2014], pg. 128 |
| "A major criticism of these 'close-to-criticality' models that at physiological temperatures, the cell membrane is well above the lo-ld phase boundary, and thus suggestions that coupling to cortical actin suppresses phase segregation, simply do not work." | Rao *et al.* [2014], pg. 129 |
| "Overall, it seems that the outer leaflet of the plasma membrane is in a uniform lipid phase state, and does not show Lo/Ld phase coexistence." | Sevcsik *et al.* [2015], pg. 6 |
| "The biological significance of temperature-induced lipid phase transitions in microsomes may be questioned since the body temperature of warm-blooded animals is well above those which produce phase transitions in the membrane lipids." | Zakim *et al.* [1975], pg. 343 |

| E.) | Phase transitions found to be associated with disease or illness states. | |
|---|---|---|
| | "Here, we highlight recent findings showing that age-related neurodegenerative diseases are linked to aberrant phase transitions in neurons." | Alberti *et al.* [2016], pg. 959 |
| | "It was observed that in T1DM the fluidification was concomitant with a decrease in AchE and Na+,K+-ATPase activities [29]." | Maulucci *et al.* [2017], pg. 10 |
| | "Keeping in mind the pathophysiological role of neuronal cholesterol, in this work, we used differential scanning calorimetry (DSC) and small angle X-ray scattering (SAXS) to explore thermotropic phase behavior and organization (thickness) of hippocampal membranes under conditions of varying cholesterol content.  Our results show that the apparent phase transition temperature of hippocampal membranes displays characteristic linear dependence on membrane cholesterol content... Defective cholesterol metabolism in the brain therefore results in a number of neurological disorders [14]." | Reddy *et al.* [2016], pg. 2611 |